# Interplay between superconducting fluctuations and weak localization in disordered TiN thin films

Sachin Yadav, [1,2] Bikash Gajar,[1,2] R. P. Aloysius,[1,2] and Sangeeta Sahoo[1,2,]*

[1]CSIR-National Physical Laboratory, Dr. K.S. Krishnan Marg, New Delhi-110012, India

[2]Academy of Scientific and Innovative Research (AcSIR), Ghaziabad- 201002, India

*Correspondences should be addressed to S. S. (Email: sahoos@nplindia.org)

# Abstract

The interplay between superconducting fluctuations (SFs) and weak localization (WL) has been probed by temperature dependent resistance [$R(T)$] and magnetoresistance (MR) measurements in two-dimensional disordered superconducting TiN thin films. Within a narrow band of temperature above the transition temperature Tc, the coexistence of SFs-mediated positive MR and WL-led negative MR in different range of magnetic field, as well as a crossover from positive to negative MR with increasing temperature are reported here. The crossover temperature coincides with a characteristic temperature ($T_{max}$) at which a resistance peak appears in the zero-field $R(T)$. The resistance peak and the associated magnetoresistance anomalies are addressed by using the quantum corrections to the conductivity (QCC) theory. We show that WL can be accounted for the observed negative MR. By introducing individual coefficients to both SFs and WL contributions, the dominance of one over the other is monitored with respect to temperature. It is observed that just above the $T_c$, SF dominates and with increasing temperature, the contributions from the both become comparable and finally, at $T_{max}$, WL takes over completely. The presented approach may be adopted to compare various quantum contributions in two-dimensional superconductors particularly in the regime where both SFs and WL are pronounced.

**Introduction**

Two-dimensional superconductivity in disordered superconductors has been one of the most investigated fields in the last decades as it provides a direct, rather simple and also very efficient platform to probe and execute many interesting quantum phenomena such as quantum phase transition,[1, 2] quantum criticality,[3, 4] quantum phase fluctuations,[5-7] superconducting fluctuations,[8] localization[9] among others. Of particular interest, superconducting fluctuations (SFs) above the critical temperature ($T_c$) play a crucial role in determining the transport properties of the system in weak-disorder regime and hence fluctuation phenomena have been extensively studied in disordered superconductors mainly by means of temperature and magnetic field dependent resistivity measurements.[10] Above $T_c$, SF consists of two main contributions, namely, the Aslamazov-Larkin (AL)[11] and Maki-Thompson (MT)[12, 13] fluctuations, among which the former is dominant at temperature close to $T_c$ and the latter can contribute significantly up to a temperature far from $T_c$. In addition to SFs, another quantum phenomenon, namely, weak localization plays also very important role in controlling the transport mechanism through disordered superconductors. It has been shown that the MT contribution from SFs and the WL show similar field dependence but with opposite sign.[10] Therefore, it is interesting to study and experimentally probe the interplay between these two opposite and competing quantum phenomena in any suitable disordered superconductors particularly in the regime where both of these contributions are significant and comparable.[14, 15]

Here, we have selected two-dimensional disordered TiN thin film as the choice of material which has already shown its promises in the execution of superconductor insulator quantum phase transition (SIT), [3, 16, 17] quantum criticality,[4] quantum phase slips[18] & phase fluctuations,[6] Berezinskii-Kosterlitz-Thouless (BKT) physics,[8, 19] quantum interference,[20] quantum Griffiths singularity (QGS) [21] and so on. Recently, we have demonstrated a novel substrate mediated nitridation technique [22-24] by which the disordered TiN films were grown for the present study.

We present here low temperature electrical transport measurements by means of temperature and magnetic field dependent resistance measurements, $R(T)$ and $R(B)$, respectively. While zero-field cooling, we observe an upturn in the $R(T)$ just before the superconducting (SC) transition which leads to a resistance maximum/peak resembling a resistive reentrant state.[25, 26] The corresponding peak amplitude in resistance is ~ 0.4% of its normal state resistance ($R_N$). Further, isothermal $R(B)$ measurements have been carried out in perpendicular field with the emphasis on the reentrant region and a similar type of resistance peak ($R_{max}$) appears with resistance higher than $R_N$. This anomalous peak in the $R(B)$ leads to negative magnetoresistance (MR) which appears at temperature above the $T_c$. The change in resistance ($\delta R$) between $R_{max}$ and $R_N$ varies with temperature in a similar fashion as that of zero-field $R(T)$. With increasing temperature, the peak gets stronger in amplitude and shifts towards lower field for the temperature window $T_c < T < T_{max}$ where both positive and negative MR coexist in different range of magnetic field. And finally the peak reaches to its maximum at $T_{max}$ where a total crossover to negative MR occurs. Further increasing temperature for $T > T_{max}$, the peak amplitude gets suppressed.

Appearance of resistance peak in $R(T)$ and negative MR have been observed in disordered superconducting thin films of $a$:InO,[27] TiN,[3, 26, 28] Pb,[29] Al-Ge,[30] NbTiN[31] etc.[32] Generally, for strongly disordered superconducting thin films at temperature far below the $T_c$ and at high field, negative MR appears at the vicinity of the magnetic field induced SIT due to strong phase fluctuations.[33] At temperature below $T_c$, negative MR around zero field also appeared for amorphous InO nanowires[34] and crystalline $Mo_2C$ flakes.[35] Most of these cases, the granularity is explained as the origin of the observed NMR where the Josephson coupling strength between individual superconducting grains plays a very crucial role on the overall transport mechanism.[30, 36]

However, for moderately disordered superconductor (with $R_N$ much lower than the quantum resistance), above $T_c$, amplitude fluctuation for the superconducting order parameter and the quantum interference originated by disorder mediated coherent scattering of quasiparticles take the lead role for the transport. Here, mainly we consider the transport at high temperature ($T > T_c$) and for a moderate magnetic field comparable to the critical field ($B_{C2}$) where quantum contributions from SFs and WL to magnetoconductivity (MC) plays the main role. Here, SFs contribute to the positive MR as resistance increases with increasing magnetic field by field-induced pair-breaking of superconducting Cooper pairs. On the other hand, contribution from WL can explain the observed negative MR as resistance decreases when magnetic field destroys the phase coherence of self-intersecting paths participating into the quantum interference. While considering the MR in the light of quantum corrections to conductivity (QCC) theory, we have introduced each dominating contributions with individual coefficients that reflect the strength of a particular mechanism. By comparing the coefficients, we have shown that SF and WL contributions can be accounted for the observed positive and negative MR, respectively. Further, we have shown that these two quantum phenomena compete with each other and the interplay between them along with their respective dominance at certain range of temperature leads to (i) SF dominating, (ii) simultaneous SF and WL governing and (iii) WL leading regimes.

## Experimental

TiN thin films were grown on undoped Si (100) substrate covered with a $Si_3N_4$ dielectric spacer layer (thickness ~ 80 nm) which was grown by low pressure chemical vapor deposition (LPCVD) technique. First, Ti film was deposited on the substrate by dc magnetron sputtering using a Ti target (99.995% purity) in the presence of high purity Ar (99.9999%) gas. Sputtering of Ti was performed with a base pressure less than 1.5 x $10^{-7}$ Torr. Finally, the sputtered sample was transferred *in situ* to an UHV chamber attached to the sputtering chamber for annealing. Ti films on $Si_3N_4$/Si (100) were annealed at ~ 750 ˚C for 2 hrs at a pressure less than 2 x $10^{-7}$ Torr. Nitridation of Ti film was done by high temperature annealing under high vacuum and the details of this substrate mediated nitridation process has been reported elsewhere.[22-24, 37] The film thickness was determined by using the optimized rate for the Ti films before and after the annealing process by the depth profile measurements using atomic force microscopy (AFM). Further confirmation of the thickness for the sample SS1 (after annealing in the nitride phase) is presented in Fig. S7 in the Supplementary Material. Here, we used the deposition rate of 5 nm per minute for the as grown Ti film which eventually converts into the rate of ~ 4nm/minute after the annealing process.

For the transport measurements, we have patterned the thin films into Hall bar geometry by using shadow mask made of stainless steel. We have used a complimentary separate mask to make the contact leads for voltage and currents probes. The device geometry for a representative device is shown in the inset of Fig. 1(a). The contact leads were made of Au (80-100 nm)/Ti (5 nm) deposited by dc magnetron sputtering. Transport measurements were carried out in dilution refrigerator by Oxford Instruments with base temperature 20 mK and equipped with superconducting magnet for magnetic field up to 14 T. For recording temperature dependent resistance [$R(T)$] measurements, the conventional 4-probe configuration was adopted by using standard Lock-In technique with 100 nA excitation at 17 Hz frequency. Here, model 7265 from Signal Recovery is used as the Lock-in amplifier. Further, a low noise voltage preamplifier (Signal Recovery 5113) was used to record the voltage signal. We have used Keithley Delta mode setup for measuring IVCs using Keithley 6221 as the current source and Keithley 2182A as the nanovoltmeter.

## Results

The $R(T)$ measurements for a disordered TiN thin film sample SS1 is shown in Fig. 1(a). The resistance, as shown by the black open circles, corresponds to sheet resistance $R_S$. The measurements have been carried out in four-probe geometry as shown in the left inset of Fig. 1(a). The experimental $R_S(T)$ is fitted with the QCC theory (the blue solid curve).

A closer view of the $R_S(T)$ just before the transition from normal (NM) to SC state is shown in the right inset of Fig.1(a) where a resistance peak of width ~ 0.2 K and amplitude of about 0.4% of $R_N$ is observed.

### *Quantum corrections to the conductivity (QCC):*

The blue curve is the fit obtained by considering the contributions from all the quantum corrections to the conductivity that include weak localization (WL) caused by the quantum interference [$\Delta G^{WL}(T)$], and the electron-electron interaction (EEI) present in the disordered system[$\Delta G^{EEI}(T)$].[38] The total conductivity (G) is then obtained by,

$$G = G_0 + \Delta G = G_0 + \Delta G^{WL} + \Delta G^{EEI} \qquad (1)$$

Where, $G_0$ is the classical Drude conductivity and $\Delta G$ is the total quantum corrections to the conductivity. The contribution from $\Delta G^{EEI}$ in disordered superconducting materials comprises of electron-electron interactions in two types of channels; one is from the diffusive single particle channel (ID) and the other one is from the channel of Cooper pairs commonly known as superconducting fluctuations (SFs). The SFs include Aslamazov-Larkin [$\Delta G^{AL}(T)$], Maki-Thompson [$\Delta G^{MT}(T)$], and the suppression in density of states [$\Delta G^{DOS}(T)$] due to cooper pair formation.[38] Therefore, the total conductivity becomes as,

$$G = G_0 + \Delta G^{WL} + \Delta G^{ID} + \Delta G^{AL} + \Delta G^{MT} + \Delta G^{DOS} \qquad (2)$$

For two-dimensional superconductors, the first two correction terms in Equation (2) vary logarithmically with temperature and they appear with a universal constant $G_{00} = e^2/(2\pi^2\hbar)$ as,[38]

$$\frac{\Delta G^{WL}(T) + \Delta G^{ID}(T)}{G_{00}} = A \ln\left[\frac{k_B T \tau}{\hbar}\right] \qquad (3)$$

Where A is a proportional constant and $\tau$ is the electron mean free time. The other terms from the SFs are,

$$\frac{\Delta G^{AL}(T)}{G_{00}} = \frac{\pi^2}{8}\left[\ln\left(\frac{T}{T_c}\right)\right]^{-1}, \qquad (4)$$

$$\frac{\Delta G^{MT}(T)}{G_{00}} = \beta(T/T_c) \ln\left[\frac{k_B T \tau_\phi}{\hbar}\right] \qquad (5)$$

$$\frac{\Delta G^{DOS}(T)}{G_{00}} = \ln\left[\frac{\ln(T_c/T)}{\ln(k_B T_c \tau/\hbar)}\right], \qquad (6)$$

Here, in MT contribution, $\tau_\phi$ introduces phase breaking processes due to mainly inelastic scattering (as spin-flip scattering can be ignored here)[39, 40] and $\beta(T/T_c)$ relates to the strength function characterizing electron-electron interaction which has been introduced by Larkin.[41] The experimental data (the open circles) obtained from the $R(T)$ measurements in Fig. 1(a) is fitted by the sum of all the aforementioned quantum corrections using the following formula,[38]

$$R(T) = \frac{1}{(\Delta G^{WL}(T) + \Delta G^{ID}(T) + \Delta G^{AL}(T) + \Delta G^{DOS}(T) + \Delta G^{MT}(T)) + 1/R_S(T=7\ K)}, \qquad (7)$$

The blue solid curve represents the best fit using Equation (7). The individual contributions as stated by the Equations (3)-(6) are shown in Fig. S2 in the Supplementary Material (SM). The coefficient '$A$' and the transport scattering time $\tau$ are obtained from the (WL + ID) fit using Equation (3) and the best-fit values are $A=2.8 \pm 0.35$ and $\tau = 2.0 \pm 1.73$ fs, in agreement with the reported values for TiN.[33, 42-44] From the MT fit using Equation (5), we obtain the critical temperature $T_c =2.45 \pm 0.003$ K and the phase relaxation time $\tau_\phi = 19.1 \pm 7.72$ ns. In order for WL to occur, the quasiparticles in the diffusion channel undergo several scattering events while maintaining their phase coherence. Therefore, the phase coherence time should be much higher than the electron mean-free scattering time and hence, From the fit, the ratio $\tau/\tau_\phi \ll 1$ satisfies the applicability condition for 2D WL theory.[45, 46]

A log-log scale presentation of the current-voltage characteristics (IVCs) at zero magnetic field for the sample is displayed in Fig. 1(b), where IVCs are observed to follow a power law ($V \propto I^\alpha$) relation shown by the dashed cyan lines. The values of the exponent $\alpha$ from the power law fits are given by the slopes of these lines and $\alpha$ determines the nature of the IVCs. For example, $\alpha = 1$ corresponds to Ohmic behavior which takes place at temperature 2.75 K and above. Further, $\alpha = 3$ corresponds to BKT transition which occurs at 1.5 K. In the inset of Fig. 1(b), we have plotted $\alpha$ values with respect to temperature and the value of $\alpha$ approaches to 3 at 1.5 K which is marked as the $T_{BKT}$ as shown by the dotted violet lines. Moreover, the zero field $R_S(T)$ is also fitted with Halperin-Nelson (HN) formula[47] as shown in Fig. S3 in the SM and yields the BKT temperature at 1.51 K. Hence, the BKT transition itself shows the 2D nature of the film where the phase fluctuation of the order parameter may occur in the form of phase slip lines.[24, 37, 48, 49]

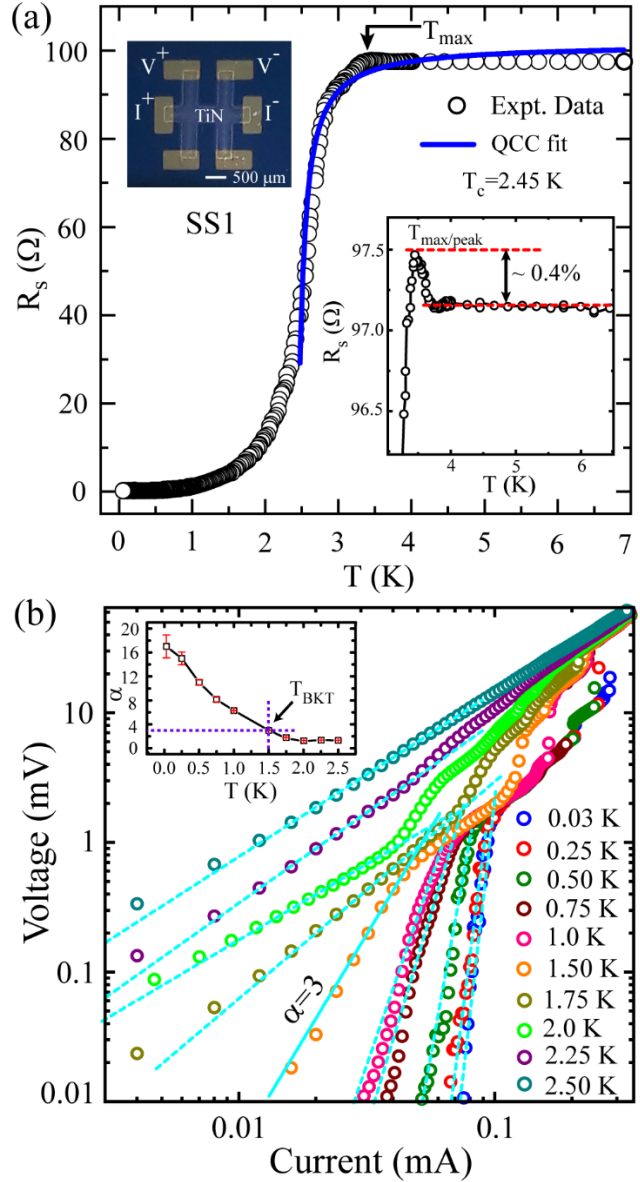

**Fig. 1:** *Transport characteristics for the sample SS1 of thickness (7±1) nm with no external magnetic field.*
*(a) Sheet resistance vs. temperature of the sample. The solid blue curves represent the QCC fit [Equation (8)] and provides the $T_c$. Insets: (Left) Device geometry with current and voltage terminals; (Right) appearance of a resistance peak in an enlarged view. (b) A log-log scale presentation of current-voltage characteristics (IVCs) with power law fittings $V = I^\alpha$ shown by the cyan dotted lines. Inset: The dependence of $\alpha$ on temperature obtained from the fit in the main panel. The temperature at the crossing point of the violet dotted lines corresponds to $\alpha = 3$, and it is the $T_{BKT} = 1.5$ K. The details of all the fittings presented in this figure are explained in the text.*

***Anomalous/negative magnetoresistance and its evolution with temperature:***

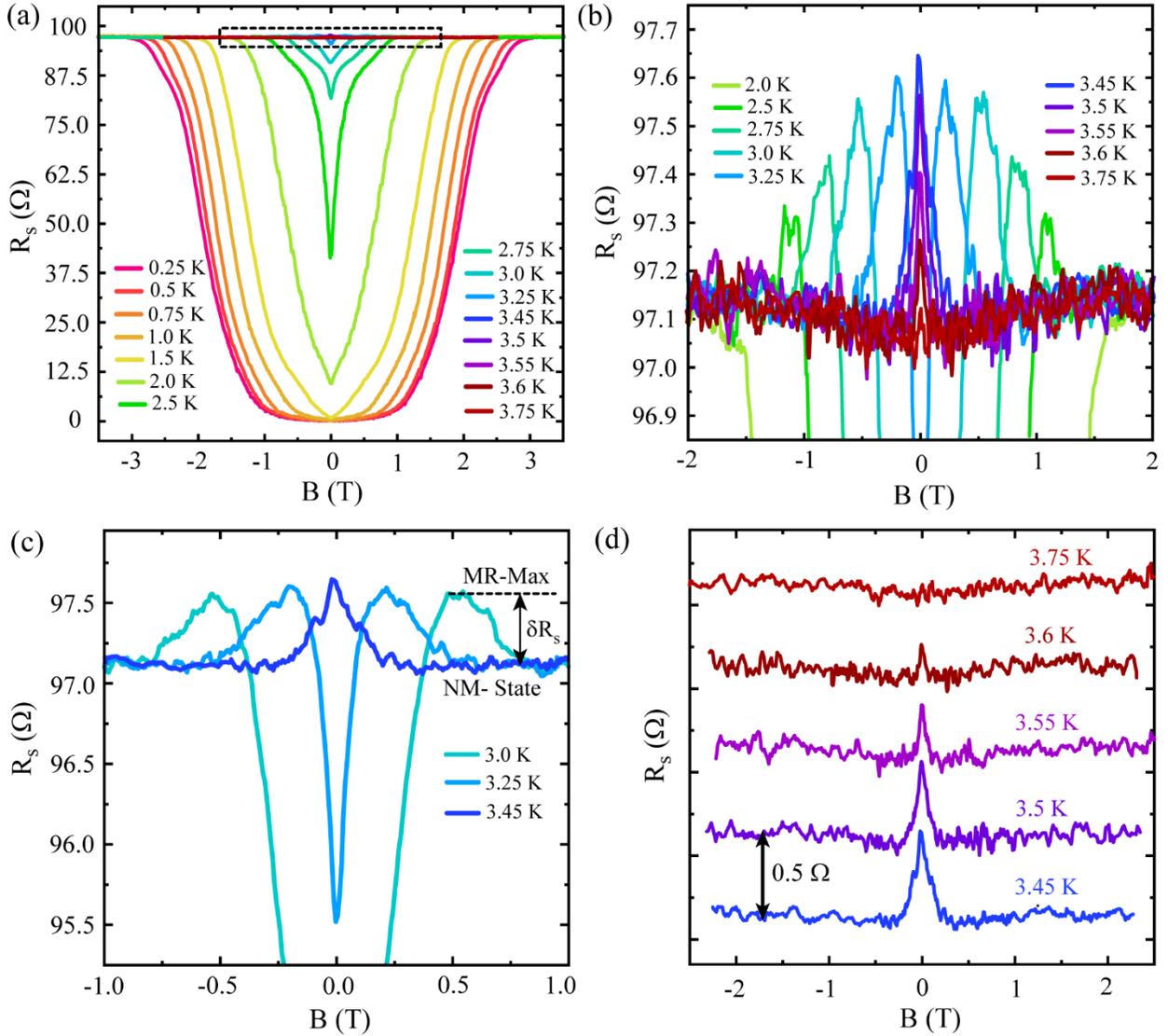

***Fig. 2:*** *Magnetic field dependent resistance $R_S(B)$ measurements for the field applied perpendicular to the sample plane. (a) $R_S(B)$ isotherms measured at various temperature from 250 mK to 3.75 K which is more than 1.5 $T_c$ ($T_c$ = 2.45 K). (b) A magnified view of a selected portion of the $R_S(B)$ close to the normal state as bounded by the dashed rectangular region shown in (a). (c) A set of three representative $R_S(B)$ isotherms showing the presence of both positive and negative MR along with their evolution with temperature and finally merging into a resistance peak around zero field indicating a total negative magnetoresistance (MR). (d) The evolution of the zero-field resistance peak with further increasing temperature. Here the $R_S(B)$ curves are shifted vertically for clarity and it is evident that the peak amplitude decreases with increasing temperature supporting the weak localization (WL) mechanism.*

The $R(B)$ measurements for the sample are carried out under the field applied perpendicular to the sample plane and the corresponding $R_S(B)$ isotherms are presented in Fig. 2. The temperature was varied from 250 mK to 3.75 K and no detectable MR was observed above 3.75 K. At temperature far below the $T_c$ ($T_c$ =2.45 K), the $R_S(B)$ curves initially show a zero-resistance state up to a certain characteristic magnetic field (the lower critical field $B_{C1}$). With further increasing field, resistance starts to show up due to field induced vortex motion and it increases sharply with the increasing field leading to a strong positive magnetoresistance. Finally, at the upper critical field ($B_{C2}$), resistance starts to merge onto the normal state.

At temperature just above $T_c$, the $R_S(B)$ isotherms show two distinct segments of positive MR with different slopes (d$R_S$/d$B$) before transiting to the normal state. The first segment reveals a very sharp positive MR cusp at low field which then transits into a much wider wing at higher field. With increasing temperature, the magnitude of the low field segment decreases and finally merges onto the wider branch at about 3.0 K. Interestingly, the wider branch does not simply blend with the normal state but before reaching to the normal state the resistance goes above the $R_N$ and then comes back to align with the normal state, hence, the $R_S(B)$ features a peak type of structure above the normal state. The region bounded by the dotted rectangle in Fig. 2(a) accommodates the aforesaid anomalous resistance peak and the region is highlighted in Fig. 2(b) which clearly shows the appearance of a resistance peak above the normal state as well as their evolution with temperature. With increasing temperature, the peak appears at lower field but with higher amplitude and finally at certain temperature ($T_{max}$) peak appears at zero field with the highest amplitude. Further increasing temperature, the peak amplitude reduces. For a better understanding of the evolution of the resistance peak appearing in the $R_S(B)$ measurements, we have displayed three representative $R_S(B)$ isotherms measured at 3.0 K, 3.25 K and 3.45 K, respectively, in Fig. 2(c), which eventually shows that with increasing temperature the resistance peak moves towards the zero-field. The peak resistance is marked as the $R_{Max}$ which differs from the normal state resistance by $\delta R_S$ as shown in Fig. 2(c). At 3.45 K with maximum $\delta R_S$, the peak appears at zero-field with only negative magnetoresistance (MR). The similar trend is observed up to 3.6 K as presented in Fig. 2(d) where the isotherms are shifted vertically for clarity. We observe that the peak amplitude for the negative MR decreases with increasing temperature and at 3.75 K, almost no detectable MR is observed.

Further, we have compared the temperature variation of $\delta R_S$ with and the zero field $R_S(T)$ in Fig. S4 in the SM. Where, only a narrow window in the temperature, mostly confined by the transition region, offers non-zero $\delta R_S$ with the appearance of the anomalous peak structure in the $R_S(B)$ measurements. The maximum $\delta R_S$ and the resistance peak for zero-field $R_S(T)$ appear at the same temperature $T_{max}$. Further, the maximum amplitude of $\delta R_S$ is about 0.5 % of the normal state resistance which is similar to the resistance change observed in the zero-field $R_S(T)$ [the right inset of Fig. 1(a)] at the peak position. Therefore, the appearance of the anomalous peak in the $R_S(B)$ might be of the same origin to that of the resistance peak for the zero-field $R_S(T)$. Further, we have collected the magnetic field values corresponding to the normal state (NM) position and the resistance-maximum ($R_{Max}$) position for any particular temperature from the measured $R_S(B)$ isotherms and have plotted the field-temperature ($B$-$T$) dependence in Fig. 3. It is clear from Fig. 3 that for $T < 2.5$ $K$, the normal state resistance ($R_N$) is the maximum resistance $R_{Max}$. However, for $T \geq 2.5$ $K$, the NM state and the $R_{Max}$ appear at different fields and the corresponding $B$-$T$ characteristics bifurcate. At $T \geq 3.45$ K, the $R_{Max}$ appears at zero-field before it disappears at $T$=3.75 $K$. Therefore, the peak structure in $R_S(B)$ isotherms spans over the temperature range $2.5 \leq T < 3.75$ $K$.

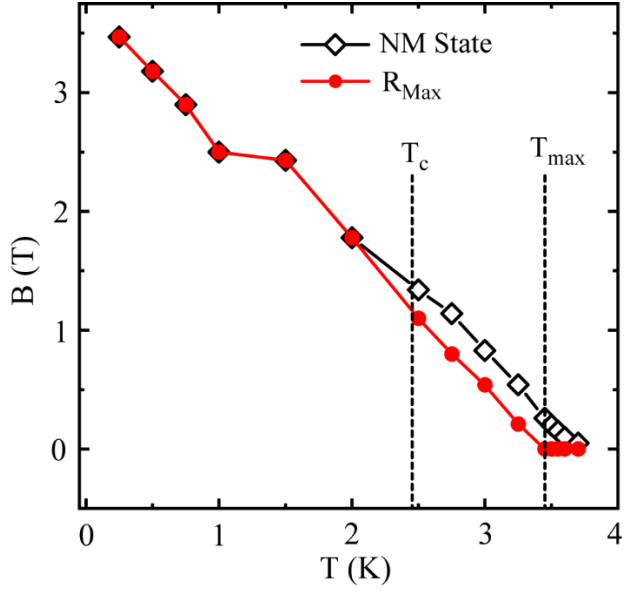

*Fig. 3: B-T dependence for the NM state and $R_{Max}$ obtained from $R_S(B)$ sweep at particular temperature. The positions of the $T_c$ and $T_{max}$ obtained from zero-field $R_S(T)$ are shown by the vertical dashed lines on the temperature axis.*

An anomalous peak, leading to negative MR with maximum amplitude of about 0.5% of $R_N$, appears too in the $R_S(B)$ measurements. For the temperature range $T_c<T<1.4T_c$, the $R_S(B)$ contains both positive and negative MR and for the temperature interval $1.4T_c<T<1.5T_c$, there is only negative MR which gets suppressed by increasing temperature. Above $1.5T_c$, almost no detectable MR is observed. Generally, weak localization is considered as one of the reasons behind the observed upturn in the $R_S(T)$ and/or the negative MR in the $R_S(B)$.[50-52] The negative MR with an amplitude of about 0.5% and its suppression with increasing temperature in the temperature region where only negative MR is observed strongly indicate about the WL to play the role for the observed negative MR.[46] However, in the temperature region where both positive and negative MR coexist in different field range, the amplitude of negative MR increases with temperature which is in contradiction with the WL mechanism.[34, 46] Here, it should be also noted, in addition to the negative MR, there is a strong influence of SF mediated positive MR which decreases with temperature and the negative MR starts to increase. Hence, if the negative MR originated from WL, due to the influence of SF in the region where both SF and WL strongly interact with each other, the amplitude of WL increases with increasing temperature till the SF gets completely suppressed.[33, 39, 53]

**Quantum contributions to the magnetoconductivity (MC) and the origin of the negative MR:**

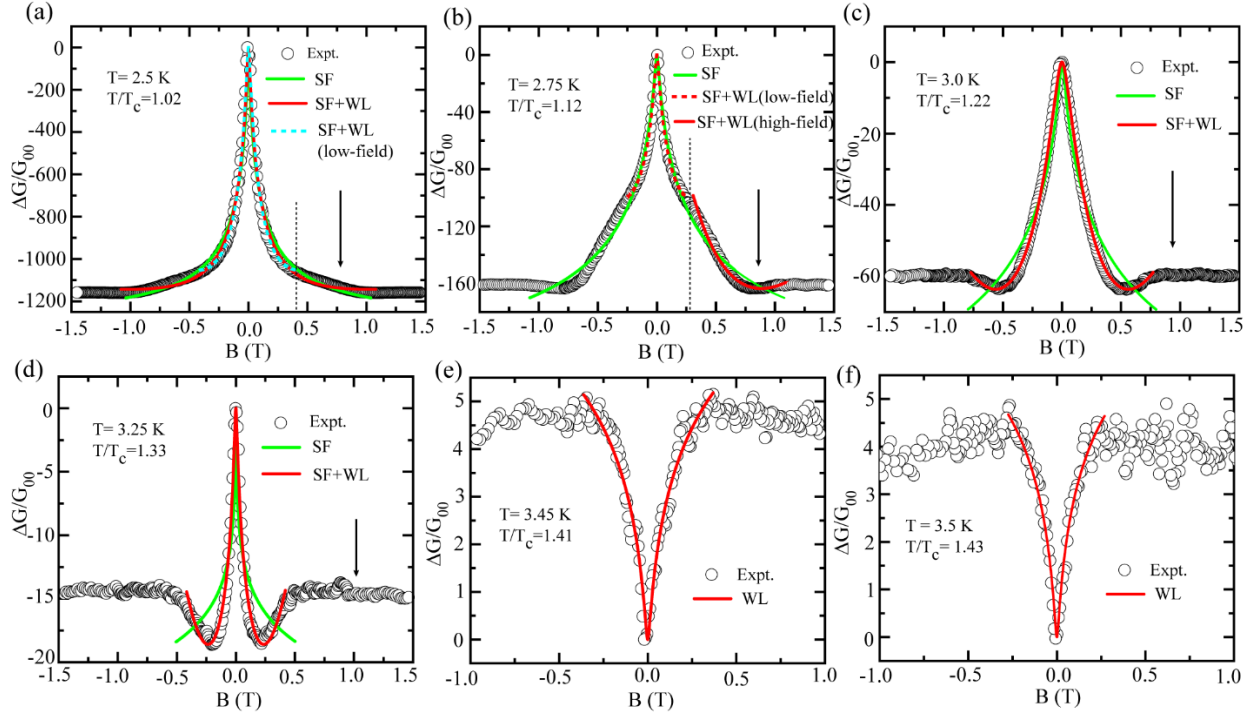

**Fig. 4:** *Presentation of a set of magnetoconductance [G(B)=1/R$_S$(B)] isotherms for the temperature regime where the anomalous peak and the consequent negative MR is observed. The combined effect of superconducting fluctuations and weak localization is considered for analyzing the experimental results and the details are described in the text. The magnetoconductance data is fitted by combining the SF and WL effects (the red solid curve) and also with only SF (the green solid curve) which is mostly dominated by the MT contribution. (a-d) The experimental data is fitted by the superconducting fluctuations with (the red solid curve) and without (the green solid curve) the WL terms. (e) & (f) Weak localization alone (with negligible contribution from SFs) can satisfactorily fit the data consisting of only the negative magnetoresistance from the anomalous resistance peak appearing at zero-field. The cyan dashed curve in (a) represents the total contribution from SF and WL for the field region up to which the fit exactly follows the experimental data. The dashed red curve in (b) represents the total fit from the combination of SF and WL at the low field region where the fit follows the experimental data completely. The vertical dotted lines in (a) & (b) divide the low field and high field region to distinguish the hump type of structures appearing in the high field region. The details are explained in the text. The vertical solid arrows in (a-d) indicate the characteristic field restricting the validity of MT contribution.*

In order to understand the mechanism behind the evolution of magnetoresistance with temperature for $T>T_c$ as observed by the presence of both positive and negative MR, we have plotted the corresponding field dependent conductivity, $G=1/R_S$ with $R_S$ as the sheet resistance, in Fig. 4. Generally, in 2D superconductors, at $T>T_c$, quantum corrections to the conductivity (QCC) through weak localization & superconducting fluctuations dominate and with the application of magnetic field superconducting fluctuations get reduced, hence resistance increases and eventually, positive MR or negative MC is observed. On the other hand, WL gets strongly suppressed by the application of field and hence, resistance decreases and positive MC or negative MR appears. However, sample's dimensionality, as determined by the relevant characteristic length scales, is very important as superconductors in 2D are the best suited candidates for the aforesaid quantum contributions to play prominent role. Here, the zero temperature superconducting coherence length $\xi(0)$ (~ 9 nm) (Fig. S1 in the SM) and the thermal coherence length, $L_T = \sqrt{2\pi\hbar D/(k_B T)}$, at normal state (~ 18 nm at 10 K) exceed the film thickness $d$ which is ~ 7 nm. Further, the dephasing or phase coherence length as obtained later from the MC analysis at T = 3.25 K is ~ 223 nm which is much more than the film thickness. Therefore, the 2D analysis related to the quantum correction to the magneto-conductivity may be applied to the present set of MR data.

Near the $T_c$, contributions to the conductivity for a superconductor arise mainly from the phase and amplitude fluctuations of the superconducting order parameter, the normal-state quasiparticles and the vortex motion. Above $T_c$ but close to the transition, an enhancement of conductivity occurs due to shunting with a parallel conductive channel formed by the fluctuating Cooper pairs. The related contribution is known as Aslamazov-Larkin (AL) contribution as it was first calculated by Aslamazov and Larkin (AL).[11] The contribution from the coherent scattering of quasiparticles that are generated by the fluctuating broken Cooper pairs before losing their phase coherence is known as Maki-Thompson (MT) contribution.[12, 13] These two contributions are main SF contributions to the conductivity above $T_c$. As both the AL and MT contributions originate from the fluctuating Cooper pairs, they are sensitive to the applied magnetic field and therefore they lead to two dominant SF contributions to the MC of the system. AL dominates near $T_c$ and MT can be significant even far from $T_c$.[54] Above $T_c$, quantum correction due to WL contributes significantly to the MC as it is very sensitive to the applied magnetic field [46]. We discuss here the afore-said three main quantum contributions, viz., AL, MT and WL, to understand the observed MC and its evolution with the temperature.

The field dependent electrical conductance can be expressed as,

$$G_{xx}(B) = G^n + \Delta G^{SF}(B,T) + \Delta G^{WL}(B,T) \qquad (8)$$

Where the first term represents the Drude's conductance, the second term originates from the superconducting fluctuation and the final term is the localization term which arises due to the disorder induced quantum interference for the normal electrons. Considering only the quantum contributions and ignoring the classical counterpart, the magnetoconductivity is given by,

$$\Delta G_{xx}(B) = \Delta G^{SF}(B,T) + \Delta G^{WL}(B,T) \qquad (9)$$

First, we consider the contribution from WL to the MC which gets suppressed under external magnetic field. Generally, several characteristic time scales such as elastic scattering time ($\tau$), magnetic impurity scattering time ($\tau_{imp}$), spin-orbit scattering time ($\tau_{so}$), dephasing time ($\tau_\phi$) etc. are involved in the WL mechanism and the corresponding characteristic fields ($B_x$) relate with the characteristic time ($\tau_x$) as,

$B_x = \frac{\hbar}{4eD\tau_x}$. The elastic scattering time $\tau$ as obtained from the QCC analysis is of the order of ~ $10^{-15}$ s which is in the similar range with the reported values of ~ $10^{-15}$-$10^{-16}$ s for TiN.[33, 42-44] The corresponding characteristic field $B_e$ is of the order ~ $10^3$ -$10^4$ T, which is much more than the other characteristic fields and it is few orders more than the experimentally achievable field. Therefore, we will not consider the case related to $B_e$. Further, in TiN, the scattering related to spin-orbit interaction can be ignored. As it is highly unlikely to have magnetic impurities in our sample, we only consider $B_\phi$ for determining the WL contribution. The WL contribution to the MC is positive and can be expressed using the model by Hikami, Larkin, and Nagaoka (HLN),[55]

$$\Delta G^{WL}(B,T) = N.\frac{e^2}{2\pi^2\hbar}.Y\left(\frac{B}{B_\phi}\right), \hspace{3cm} (10)$$

with $Y(x) = \ln(x) + \psi\left(\frac{1}{2} + \frac{1}{x}\right)$ and $B_\phi = \frac{\hbar}{4eD\tau_\phi}$, $D = \frac{\pi}{2\gamma}\frac{k_B T_c}{eB_{c2}(0)}$, $\gamma$=1.78, $G_{00} = \frac{e^2}{2\pi^2\hbar}$ (10a)

where, $\psi(x)$ is the digamma function and $\tau_\phi$ is the dephasing time. $D$ is the diffusion constant. Here, the coefficient $N$ ($\alpha$ in the original HLN model) in Equation (10) represents the number of channels participating into the conduction process.[40, 56] Theoretically, $N = 1$ when the spin-orbit scattering and magnetic scattering are absent.[55] However, often it is considered as a free parameter[45] for the analysis of experimental MC using the HLN model and the deviation from the theoretical value for $N >$1indicates the contributions from other conducting channels and from the bulk.[56-59]

The $\Delta G^{SF}$ comprises of two main components, the AL and the MT contributions, that lead to the suppression of conductance with magnetic field. Hence, $\Delta G^{SF}$ leads to negative MC which is opposite to that of the WL contribution. The AL contributions to the MC is given by[60]

$$\Delta G^{AL}(B,T) = G_{00}\frac{\pi^2}{8ln\left(\frac{T}{T_c}\right)}\left\{8\left(\frac{B_{SF}}{B}\right)^2\left[\psi\left(\frac{1}{2} + \frac{B_{SF}}{B}\right) - \psi\left(1 + \frac{B_{SF}}{B}\right) + \frac{B}{2B_{SF}}\right] - 1\right\} \hspace{1cm} (11)$$

Where, $B_{SF}$ is the characteristic field representing the superconducting fluctuation and can be expressed by the Ginzburg-Landau relaxation time $\tau_{GL}$ as, $B_{SF} = \frac{\hbar}{4eD\tau_{GL}}$ with $\tau_{GL} = \frac{\pi\hbar}{8K_B Tln(T/T_c)}$ .

Finally, the MT contribution to the MC can be written as,

$$\Delta G^{MT}(B,T) = -G_1.\frac{e^2}{2\pi^2\hbar}\left[\psi\left(\frac{1}{2} + \frac{B_\phi}{B}\right) - \psi\left(\frac{1}{2} + \frac{B_{SF}}{B}\right) + ln\left(\frac{B_{SF}}{B_\phi}\right)\right], \hspace{2cm} (12)$$

Where,

$$G_1 = C . \beta_{LdS}(T/T_c, \delta), \; \beta_{LdS}(T/T_c, \delta) \equiv \pi^2/4(\epsilon - \delta), \; \epsilon \equiv ln\left(\frac{T}{T_c}\right), \; \delta = \frac{\pi\hbar}{8k_B T}\frac{1}{\tau_\phi} \quad (13)$$

Here, $\beta_{LdS}$ represents the effective electron-electron attraction strength for temperature close to $T_c$ at the limit $\epsilon \ll 1$ and for higher magnetic field up to $B < k_B T/4eD$[39, 61]. $\delta$ is the pair breaking or cut-off parameter[33, 39] and at higher temperature, $\beta_{LdS}(T/T_c, \delta)$ takes the form of original $\beta_L(T/T_c)$ values introduced by Larkin.[39, 61] We have introduced here an extra term/coefficient $C$ in MT contribution as presented in Equation (13) and the product of $C$ and $\beta_{LdS}(T/T_c, \delta)$, i.e., $G_1$ acts as the effective coefficient measuring the strength of the MT contribution. As MT and WL contributions have similar magnetic field dependence but with opposite sign,[10] the interplay between these two quantum contributions can be probed and studied by comparing the corresponding coefficients, i.e., $N$ from Equation (10) for WL and $G_1$ from Equation (12) for the MT contribution.

To understand the negative magnetoresistance related to the peaks appeared in the MR for the temperature range $T_c < T < 1.5T_c$, as presented in Fig. 2, we analyze the corresponding magnetoconductivity by considering the aforementioned quantum corrections. We calculate MC from the experimental field dependent longitudinal sheet resistance $R_{xx}(B)$ as:

$$\Delta G_{xx}(B) = \frac{1}{R_{xx}(B)} - \frac{1}{R_{xx}(0)} = -\frac{R_{xx}(B) - R_{xx}(0)}{R_{xx}(B).R_{xx}(0)} \quad (14)$$

We have plotted $\Delta G_{xx}(B)/G_{00}$ with $G_{00} = \frac{e^2}{2\pi^2\hbar}$ in Fig. 4 for a set of temperature points and we have analyzed the magnetoconductivity (MC) data by using aforementioned quantum contributions as expressed in Equations (9)-(13). Here, the temperature of interest extends up to only $1.5T_c$ from the $T_c$, hence, the AL and the MT contribution with $\beta_{LdS}(T/T_c, \delta)$ are considered as the SF components in addition to the WL part. In order to have a clear comparison between SF and the WL part, we have fitted the MC data with only SF given by the AL and MT contributions (the cyan solid curve) and also with the total quantum contribution by combining SF and WL parts (the red solid curve) in Fig. 4. At $T$=2.5 K ($T/T_c$=1.02), low field region fits better with the total contributions from SF and WL as shown by the red curve compared to that with only SF contribution presented by the cyan curve as shown in Fig. 4(a). To make it clearer, we have fitted only at the low field region up to the dotted vertical black line by the total contribution and the fit is shown by the dashed blue curve. The fit follows the experimental data almost perfectly. However, at the higher field at about 0.5 T and above, it is clear from Fig. 4(a) that both the fits deviate from the experimental points. Particularly in the region of *0.5 T < B < 1.0 T*, the observed hump type of structure in the experimental MC is not possible to be explained by these two curves. The arrow indicates the characteristic field which restricts the validity of the MT contribution given by Equation (12). With a little increase in temperature at $T$=2.75 K ($T/T_c$ ~1.12) in Fig. 4(b), we observe two distinct negative MC segments, separated by the dotted vertical black line, with increasing field up to ~ 0.75T at which a third segment with positive MC starts to build up. This positive MC region extends up to ~ 1 T before merging onto the normal state conductance. This third region corresponds to the negative MR which appeared above the normal state resistance in the $R_S(B)$ as shown in Fig. 2(b). At this temperature, the hump appears more prominently for the second MC segment compared to that in the previous case at 2.5 K shown in Fig 4(a). The fitting is done using all the three quantum contributions for the whole range of MC and also in two different parts corresponding to low field and high field regions. For the full MC region up to the normal state conductance, the fit presented by the cyan curve is obtained by combining AL, MT and WL with the coefficients for the latter two considered as free parameters. The coefficient of WL appears to be negligible (~ $10^{-9}$) in the least square fit and hence, the fit relates only to the SF contribution. However, the fit shown by the cyan curve does not follow the MC data for the field beyond the 1st region marked by the dotted vertical line. Further, we have fitted individually the low field region consisting of mainly the 1st negative MC segment and the high field region consisting of the other two segments of negative and positive MC till the normal state conductance. The corresponding fits are shown in red with the dashed and solid curves for the low field and the high field regions, respectively. For both of these two regions, all the three quantum contributions were considered in the respective fits. Compared to the full range fit (the cyan curve), the region wise separate fits (the red curves) follow the experimental data reasonably well, particularly fitting at the positive MC region before reaching to the normal state indicates a dominant contribution from WL.

With further increase in temperature at 3.0 K as shown in Fig. 4(c), previously observed two separate negative MC regions get merged and the MC varies smoothly up to about 0.5T at which point positive MC starts to appear. The cyan curve representing only the SF contribution clearly shows the deviation from the experimental data particularly for the field above 0.2 T. Whereas, the total quantum contributions to the MC with SF and WL as represented by the red curve follows very nicely the experimental data till the normal state MC is reached. Here, the role of WL particularly for the high field positive MC is evident from the fits shown in Fig. 4(c). Similar trend is observed for the temperature 3.25 K as shown in Fig. 4(d) which clearly displays the differences in fits with and without the WL contribution to the SF counterpart. However, the overall width gets reduced and the positive MC part appears more prominently. With respect to the amplitude and the extent in magnetic-field, the negative and positive MC become comparable at this temperature. The fit corresponding to the total quantum contribution by the combined effect of SF and WL follows well enough the experimental data points in both the negative and positive MC regions. The fitting analysis here is indicative of WL being the possible reason for the positive MC (or negative MR) that appeared before merging onto the normal state.

With further increasing temperature, at $T = 3.45$ K, we observe only positive MC which is shown in Fig. 4(e). The fit shown by the red solid curve consists of WL part only as WL alone can fit the experimental data satisfactorily. Here, it should be noted that $T = 3.45$ K is the $T_{max}$ as shown in the zero-field $R_S(T)$ in Fig. 1(a). Similar behavior of only positive MC is observed in the MC for temperature up to 3.6 K. The MC and corresponding WL fit for $T$=3.5 K is shown in Fig. 4(f). With increasing temperature, the amplitude of the MC decreases which may also indicate the WL to be the origin behind the crossover from negative to positive MC.

Moreover, a logarithmic temperature dependence for conductivity is expected from the contribution from WL in 2D.[46, 62-64] Here, for the sample SS1 presented in this manuscript, $R_S(T)$ measurements are done only up to 7K and no high temperature data is available and hence, it is difficult to show the logarithmic temperature dependence for the conductivity on the available dataset. However, $R_S(T)$ measurements on another similar sample SS3 (Table S1 in SM) in high temperature range show the logarithmic temperature dependence of conductivity. For the sample SS3, the zero-field $R_S(T)$ measurements, corresponding logarithmic temperature dependence of the conductivity and the $R_S(B)$ measurements with crossover from positive to negative MR at the peak temperature $T_{max}$ are shown in Fig. S5 in the SM. Therefore, the anomalous resistance peak and associated negative MR observed in Fig. 2 (same as the positive MC in Fig. 4) for the sample SS1 is expected to be associated with WL mechanism.

However, within a short span of temperature the aforementioned positive MC disappears and the MC curve becomes almost independent of the field variations. Also, the decrement in width for the conductance dip (resistance peak) for the temperature range 3.45 K$\leq T \leq$ 3.6 K does not relate to the characteristics of WL, as in WL mechanism, the MR width gets broadened with increasing temperature. Here, we emphasize that apart from the strong variations in the MC (MR) near the zero-field, the background normal state conductance (resistance) as shown in Fig. 4 (Fig. 2) shows a slight negative/downward (positive/upward) slope with increasing field particularly at higher temperature for T$\geq$3.45K. Though it is very difficult to extract and quantify any reasonable MC signal from the background normal state conductance, this can be attributed to MT fluctuations that generally contribute at temperatures far from the $T_c$. Hence, the strong interaction between MT and WL might lead to the suppression of the width broadening for WL contribution while increasing the temperature.[33, 39] However, further study is needed to clarify the background positive MR that might appear due to the MT fluctuation but suppressed by the presence of WL and to understand the interaction between these two mechanisms WL and MT for a wider range of temperature so that their influence on each other with respect to temperature and magnetic field can be probed in greater detail.

In order to probe the interplay between WL and MT, in Fig. 5, we have compared their relative coefficients as obtained from the fitting analysis presented in Fig. 4. The coefficients for MT (the blue spheres) and WL (the red spheres) are extracted from the best fit for maximum span of field region and their relative strength is compared in Fig. 5(a). Here, only within a narrow temperature span of $1.02T_c$ $\leq T \leq 1.33T_c$, the anomalous MR consisting of positive and negative MR in different field range and also a crossover from positive to negative MR have been observed with varying temperature. From Fig. 5(a), initially at the closest proximity of $T_c$, MT dominates over the WL mechanism and with increasing temperature, the gap between the two starts to decrease and at about 3.0 K the gap almost closes and both MT and WL become comparable. Finally, at 3.25 K, both the coefficients merge. Finally, WL takes over at 3.45 K when positive MC dominates solely with a very negligible negative MC in the background. With further increasing temperature, the gradual reduction in the WL coefficient justifies the WL mechanism to be one of the main reasons for the observed negative MR.

Further, we have calculated the phase coherence length $L_\phi$ by using $L_\phi = \sqrt{D.\tau_\phi}$, where, $\tau_\phi$ is obtained from the fitting at particular temperature and the diffusion constant $D$ is calculated from the $B_{C2}(0)$ using Equation (10a) and for this sample, $D \sim 0.673$ cm$^2$/sec. At 3.25 K, where the coefficients for both MT and WL almost match with each other, $\tau_\phi$ is obtained as $\sim 0.74$ ns and the corresponding $L_\phi$ is $\sim 223$ nm which is much more than the thickness, hence, 2-dimensionality is justified for the sample and the QCC analysis is applicable on the sample.

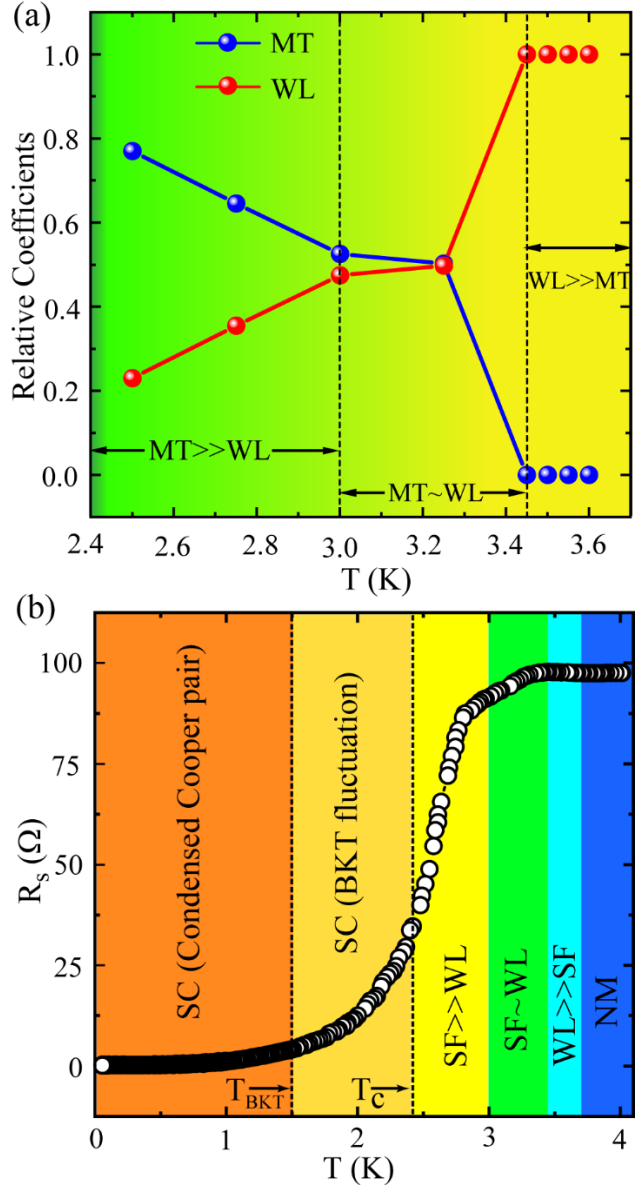

**Fig. 5:** *Comparison of the coefficients for MT and WL from the fitting in their relative strength (coefficient of each divided by the sum of the two). The coefficients are extracted from the best fit for maximum span of field region. For example, the coefficients for the fit at 2.5 K are collected from the total (SF+WL) fit presented for the low-field region up to which the fit follows the experimental data. For T= 2.75 K, the coefficients are extracted from the total fit at the high field region which covers both negative and positive MC segments. For the other temperatures, the total fit with both WL and SF is considered for the coefficients. (b) The zero-field $R_S(T)$ is divided into several regimes based on the transport analysis.*

Based on the experimental observations and also from the fitting analysis, the zero-field $R_S(T)$ is divided into several regimes that are highlighted in Fig. 5(b). First two regimes in the temperature increasing direction are formed by using the characteristic temperatures $T_{BKT}$ and the $T_c$, respectively. Below $T_{BKT}$, the system is in condensed phase with global superconductivity and is denoted as SC (condensed Cooper pair phase). The second region between $T_{BKT}$ and $T_c$ is mainly dominated by phase fluctuations. At this regime, thermally activated phase slips and movement of unbound vortex-antivortex pairs dominate and the region is marked as SC (BKT fluctuation) regime.[7, 8] The third region starts from the temperature just above the $T_c$ where the superconducting fluctuations (GL or amplitude fluctuation) and the related quantum corrections are involved. By comparing the coefficients of MT and WL from the analysis of the quantum contributions to the MC, the third region shows dominance of SF over WL. In the fourth region, both MT and WL become comparable and the region is marked accordingly. With further increasing temperature, we have observed that WL takes over completely and the related fifth region displays the supremacy of WL over MT. Finally, at temperature above 3.75 K where the MC becomes almost flat [Fig. 2(d)], the region is marked as normal state. Though in this regime, the background resistance presents a slight positive MR indicating a possible contribution from MT. However, at this regime, it is very difficult to extract a reasonable signal related to positive MR from the background noise of the measurement and hence the region may be called as the normal state (NM).

**Discussion**

The experimental magnetotransport results are analyzed by considering the AL, MT and WL contributions as the major quantum contributions to the MC. We have introduced an extra coefficient [$C$ in Equations (12)-(13)] in the expression for MT contribution in addition to the conventionally used coefficient $\beta_L(T/T_c)$ or $\beta_{LdS}(T/T_c, \delta)$.[20, 39, 61] The additional coefficient is used as free parameter in the fitting. Originally, based on the value of the ratio $T/T_c$, the coefficient $\beta_L(T/T_c)$ takes different form (Equation 13) in different range of temperature.[41, 61] For uniform fitting in the wide range of temperature, $\beta_L(T/T_c)$ or $\beta_{LdS}(T/T_c, \delta)$ is used in the literature as fitting parameter also.[10, 33] Similarly, we have conducted another check by considering the effective coefficient $G_{1=}C.\beta_{LdS}(T/T_c, \delta)$, [from Equation (13)] as a single fitting parameter and the values are coming the same as that when we use $C$ as the fitting parameter with $\beta_{LdS}(T/T_c, \delta)$ as expressed in Equation (13).

Therefore, by using an extra coefficient, we actually obtain the amount of deviation from the conventional procedure and also at the same time a better fitting for the experimental data. For the WL contribution, the coefficient is also kept as the fitting parameter so that the comparison and interplay between the SF (the sum of AL and MT contributions) and WL contributions can be understood by their respective coefficients.[61] Further, we have conducted a comparison test to justify the need of an extra coefficient in the fitting analysis by comparing the fit with and without the extra coefficient $C$ for low field as well as for the full field range containing both negative and positive MC regions for $T = 3.0$ K. The comparison is shown in Fig. S6(a) & (b) in the SM. It is evident that without the coefficient the fit does not follow the experimental data even for the low field range [Fig. S6(a)]. The MC data measured for another sample with higher resistance ($R_{Max} \sim 139$ $\Omega$/Square, Table S1 in the SM) is shown in Fig. S6(c), where the fit using the conventional theory based on fluctuation conductivity and WL [Equations (9)-(13)] with no extra coefficient ($C = 1$ and $N = 1$) follows the experimental data satisfactorily for the field range $B < k_B T/4eD$.[39, 61]

Here in this article, the present study reveals dominance of both positive MR and negative MR at any particular temperature for different field range, and the comparison of their coefficients provides an estimate about their relative contributions. Generally, in disordered superconductor above $T_c$, positive MR is most prominent and the existing formula can fit the data fairly well, however, when the contributions of both MT and WL become comparable, the present approach can be used to fit the experimental data in a better way as it was reported that deviation from the experimental data from the fit can be reduced by using $\beta_L(T/T_c)$ as free parameter.[33, 40] We have summarized all the measured samples in Table S1 in the SM where two sets are categorized based on the samples showing positive and negative MR. Apparently, it may appear that the thickness plays a major role in controlling the MR as it has been recently predicted theoretically for the thickness controlled $T_c$ in epitaxial ultrathin films.[65] However, the annealing pressure is one of the important parameters here which plays the key role in controlling the microstructural changes leading to the negative MR. For a better classification of samples based on the effects of their growth parameters on the observed MR phenomena, further study is needed.

Finally, the following points strongly indicate that WL may be the cause behind the observed NMR. (i) The amplitude of the NMR peak is very small about ~ 0.5% of the normal state resistance and generally a change in MR within 2-3% can be accounted for WL mechanism.[34, 46] (ii) The amplitude of the NMR peak decreases strongly with the temperature indicating the quantum nature of the mechanism. (iii) The experimental results related to the coexistence of positive and negative MR that appear in different magnetic field range at temperature just above $T_c$ indicates the presence of two opposite quantum phenomena that are competing with each other. Above $T_c$, quantum contribution from MT is one of the mechanisms which leads to the positive MR. Now, MT can be considered as the opposite of WL,[14] therefore, the other mechanism is most likely to be the WL which can be accounted for the observed NMR. (iv) Finally, the fitting analysis of the quantum contribution to the MC suggests and demonstrates the interplay between MT and the WL in the sample and also their evolution with the temperature.

## Conclusions

In conclusion, above the superconducting transition, a resistance peak of about 0.4% of $R_N$ has been observed in the zero-field $R_S(T)$ measurements carried out in disordered TiN thin film. Though the peak amplitude is rather small but it is distinctly present in the $R_S(B)$ measurements in the form of negative MR and its evolution with temperature follows the similar trend as that is observed in zero-field $R_S(T)$. For a narrow band of temperature, the $R_S(B)$ isotherms feature both positive and negative MR in different field range. In the analysis for quantum contribution to the MC, by introducing an extra coefficient into the conventional expression for the MT contribution we show that the modified MT expression along with the WL contribution offers a better fitting to the experimental data. By comparing the coefficients of MT and WL contributions, we have shown that these two quantum phenomena compete with each other. Based on the detailed analysis of the magnetotransport data, the zero-field $R_S(T)$ is divided into several regimes that exhibit the regions of dominance for SF and WL along with their coexistence and crossover from one to the other. Finally, the approach followed in the present study can be used in general to study the quantum corrections to the magnetoconductivity and to compare various quantum contributions in two-dimensional superconductors particularly in the weak disorder regime where contributions from both SF and WL are comparable and pronounced.

## Conflicts of interest

The authors declare no competing financial and/or non-financial interests in relation to the work described.

## Data availability

The data that represent the results in this paper and the data that support the findings of this study are available from the corresponding author upon reasonable request

## Acknowledgements

We gratefully acknowledge Dr. Sudhir Husale for the invaluable discussions and critically reviewing the manuscript. We are thankful to Mr. M. B. Chhetri for his assistance in the lab. S.Y. acknowledges the Senior Research fellowship (SRF) from UGC. B.G. acknowledges financial support from UGC-RGNF for providing research fellowship. The central facilities from CSIR-NPL are highly acknowledged. Authors acknowledge the financial support for establishing the dilution refrigerator facility at CSIR-NPL from the Department of Science and Technology (DST), Govt. of India, under the project, SR/52/pu-0003/2010(G). This work was supported by CSIR network project 'AQuaRIUS' (Project No. PSC 0110)

# Interplay between superconducting fluctuations and weak

# localization in disordered TiN thin films

*Sachin Yadav, [1,2] Bikash Gajar,[1,2] R. P. Aloysius,[1,2] and Sangeeta Sahoo[1,2,]\**

*[1]CSIR-National Physical Laboratory, Dr. KS Krishnan Marg, New Delhi-110012, India*

*[2]Academy of Scientific and Innovative Research (AcSIR), Ghaziabad- 201002, India*

*\*Correspondences should be addressed to S. S. (Email: sahoos@nplindia.org)*

***Contents:***

1. Calculation of Ginzburg- Landau (GL) coherence length ($\xi_{GL}$) for TiN sample

2. Quantum corrections to the conductivity (QCC) fit along with its individual components

3. Halperin-Nelson fit to zero field $R_S(T)$ curve to confirm the 2D nature

4. Correlation of $\delta R_S$ with the zero field $R_S(T)$ for the sample SS1

5. Zero-field $R(T)$ and $R(B,T)$ data for sample SS3

6. Comparison between MC fittings between with and without extra coefficients for WL & MT terms

7. Thickness measurement by AFM

8. Collection of TiN samples having different growth conditions

# 1. Calculation of Ginzburg- Landau (GL) coherence length (ξ<sub>GL</sub>) for TiN sample

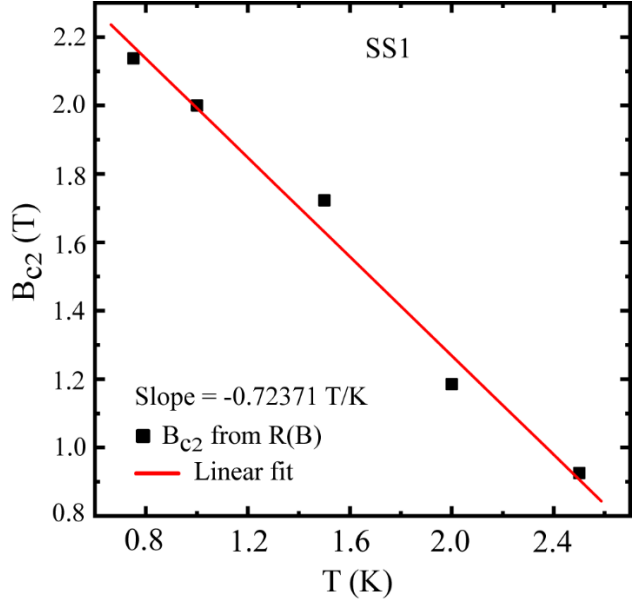

**Fig. S1:** *B-T phase diagram for the TiN sample SS1. Black squares are the data points collected from R(B) isotherms and the solid red line represents the linear fit to the experimental points which provides the slope for calculating the Ginzburg-Landau (GL) coherence length ξ<sub>GL</sub>.*

The Ginzburg-Landau (GL) coherence length $\xi_{GL}(0)$ for sample SS1 is calculated by using the formula, $\xi_{GL}(0) = \left[\frac{\phi_0}{2\pi T_c \left|\frac{dB_{c2}}{dT}\right|_{T_c}}\right]^{1/2}$, where $\phi_0$ is the flux quantum. The experimental data points are collected from the upper critical field ($B_{c2}$) values from respective *R(B)* isotherms. The extracted values from *R(B)* are fitted linearly in Fig. S1 as shown by the red line. The slope obtained from the linear fit has been used for calculating the coherence length $\xi_{GL}(0)$ for the TiN sample with $T_c$ = 2.45 K and the obtained coherence length is about ~ 9 nm.

**2. Quantum corrections to the conductivity (QCC) fit along with its individual components:**

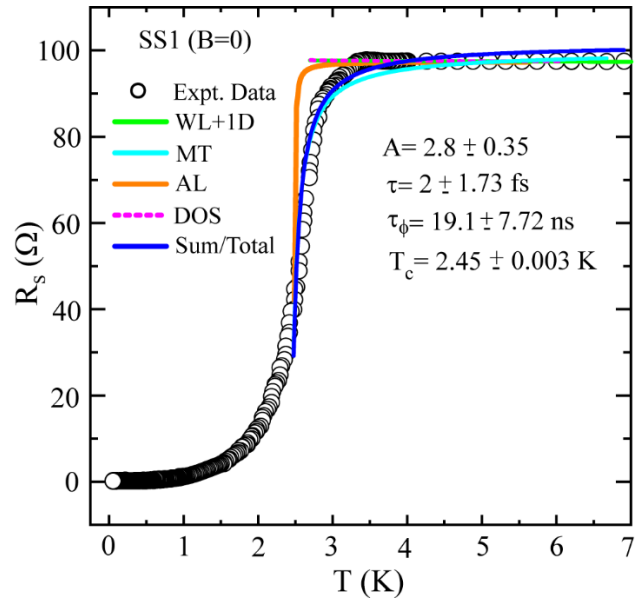

**Fig. S2:** *QCC fits to zero-field R(T) data for sample SS1 using Equations (3)-(7) from the main article that show the fits related to the individual quantum contributions and also the sum of all the contributions. Here, $T_c$ =2.45 K, obtained from MT fit.*

**3. Halperin-Nelson fit to zero field $R_S(T)$ curve to confirm the 2D nature:**

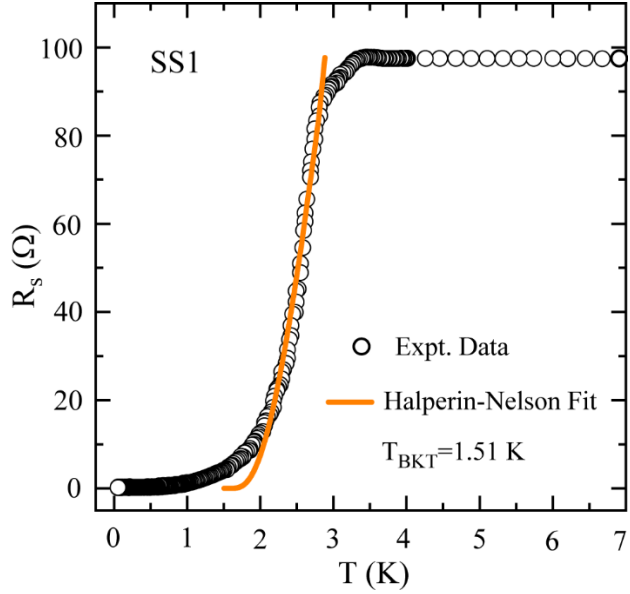

***Fig. S3:*** *Zero field $R_S(T)$ curve of the sample SS1. The solid orange curve represent the Halperin-Nelson fit [Equation (1)], where $T_{BKT}$ is obtained at 1.51 K.*

The experimental $R_S(T)$ is fitted with Halperin-Nelson (HN) formula (the orange solid curve) and given by expression, [1]

$$R_S(T) = R_0 exp\left[-b/(T - T_{BKT})^{1/2}\right], \qquad (1)$$

where, $T_{BKT}$ is the BKT transition temperature and $R_0$ and $b$ are constants. The values of the constants $R_0$ and $b$ from the fit are 4326 and 4.49, respectively and the $T_{BKT}$ = 1.51 K. However, the deviation from the experimental data at low temperature indicates that the BKT transition is suppressed at low temperature. The deviation originates from finite size effects and inhomogeneity [2,3] or it could be due to macroscopic quantum tunneling. [1]

**4. Correlation of δR$_S$ with the zero field $R_S(T)$ for the sample SS1:**

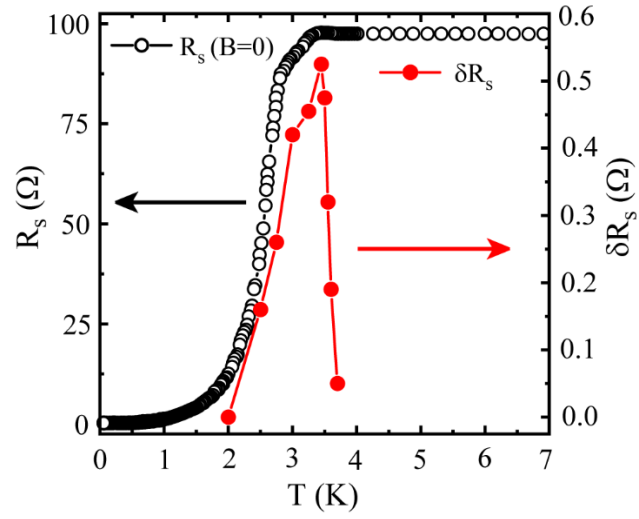

***Fig. S4:*** *Comparison of δR$_S$ with respect to the zero-field $R_S(T)$*

## 5. Zero-field *R(T)* and *R(B,T)* data for sample SS3:

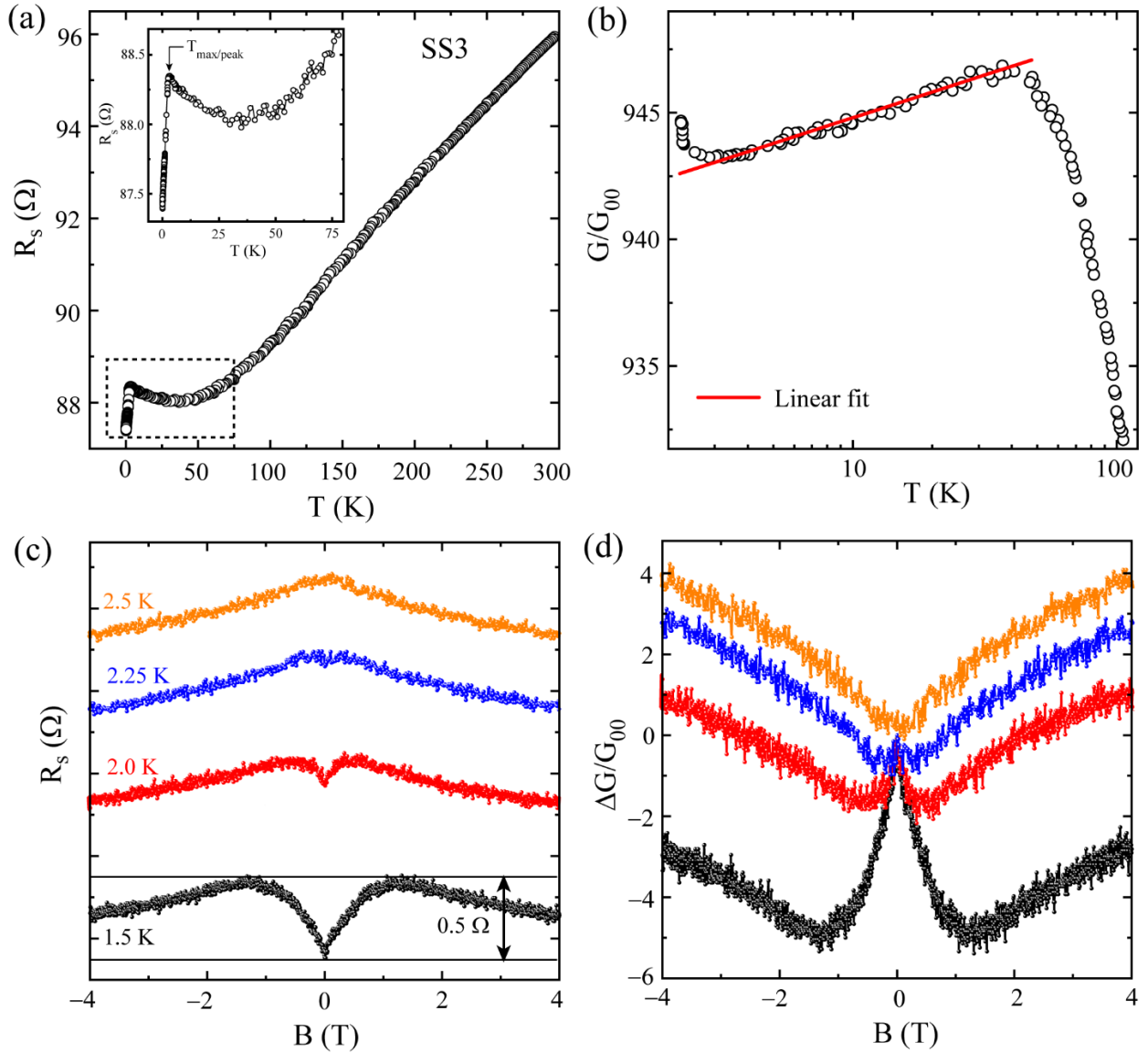

**Fig. S5:** *Zero-field R(T) & R(B,T) measurements for sample SS3. (a) Zero-field R(T) data measured from room temperature (300 K) down to 1 K. The region marked by the black dotted rectangle is highlighted in the inset which shows a resistance minimum/dip followed by a resistance maximum/peak before transiting to the superconducting state while cooling down the sample. The deviation from normal metallic behavior is evident by the prersence of the upturn in the R(T) which generally occurs due to electron elctron interaction and/or weak localization in disordered superconductors. The temperature $T_{max/peak}$ corresponding to the resistance peak/maximum is marked by the arrow in the inset of (a). (b) The logarithmic temperature dependence of R(T) in terms of dimenisonless conductance. The red line is the linear fit to the experimental data in log(T) scale. (c) A set of selective R(B) isotherms measured at different temperatures for sample SS3. The R(B) isotherms are shifted vertically for clarity to show a crossover from positive to negative magnetoresistance as T increases towards $T_{max/peak}$. (d) The corresponding magnetoconductance in unit of dimenisonless conductance showing the evolution of MC curves with temperature.*

## 6. Comparison between MC fittings between with and without extra coefficients for WL & MT terms:

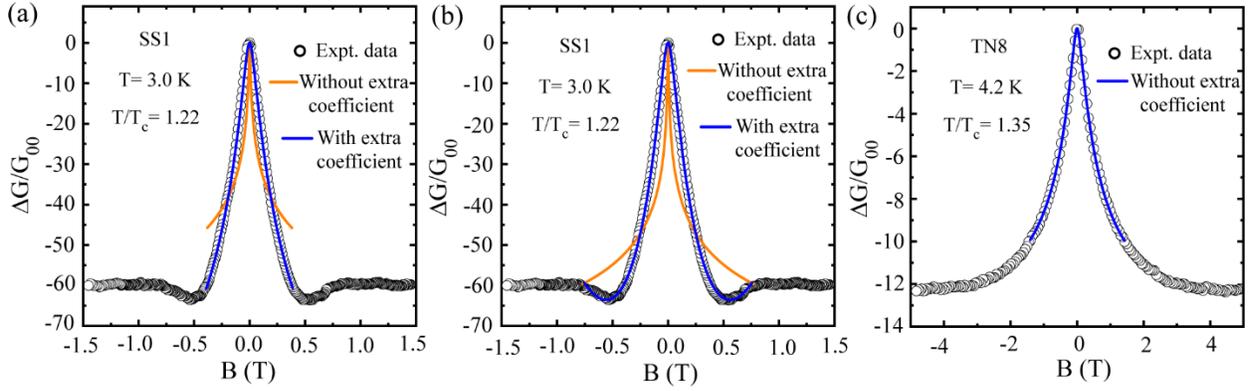

***Fig. S6:*** *Quantum corrections (SF+WL) to magnetoconductivity fits with (the blue solid curve) & without (the orange curve) extra coefficients attached to MT and WL terms for sample SS1 (a) in the low field range covering only negative MC region and (b)in the high field range containing both negative and positive MC regions. (c) The quantum corrections to magnetoconductivity fit with no extra coefficient for another sample TN8 with higher normal state resistance and with mostly negative MC. The open circles represent experimental data and the solid curves act for the fits.*

**7. Thickness measurement by AFM:**

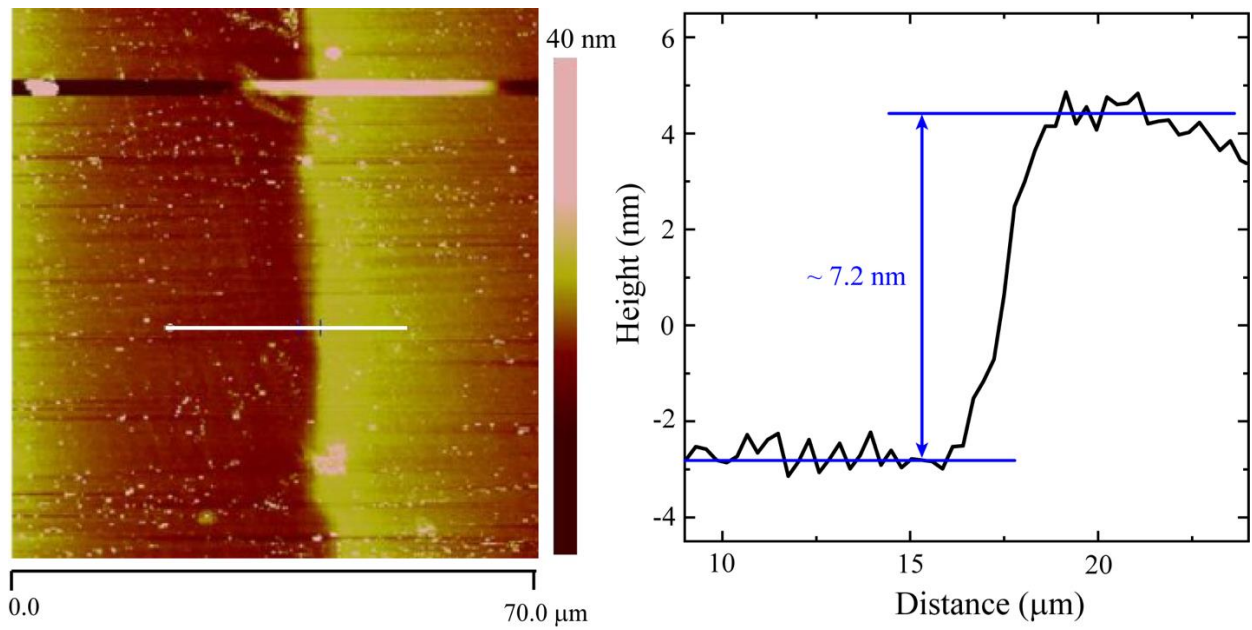

***Fig.S7:*** *(Left) AFM image of the TiN thin film sample (SS1) (after the nitridation) and (Right) the corresponding film thickness is estimated by height profile analysis.*

## 8. Collection of TiN samples having different growth conditions:

We have carried out similar measurements on many other samples that show resistance upturn in their zero-field $R(T)$ characteristics and the sample details are collected in Table S1. We observe that the upturn and the related zero-field resistance peak lead to negative magnetoresistance at the peak temperature ($T_{max}$) for the samples with normal state square resistance in the range of 85-100 $\Omega$ and the thickness in the range of 5-8 nm. On the contrary, positive magnetoresistance is observed at the same temperature $T_{max}$, corresponding to the peak temperature in zero-field $R(T)$ measurements, for the samples with higher square resistance. The second set of higher resistive samples undergo magnetic field induced SIT at higher field (data not shown here).

**Table S1:** Parameters of the TiN films measured in the course of the present study. $R_{Max}$ and $R_N^{300K}$ are in resistance per square. $T_a$ is the annealing temperature for the growth. $T_{max}$ is the temperature corresponding to the resistance peak appears in zero-field $R(T)$. $d$ corresponds to the thickness of the films. D denotes the diffusion constant for all the TiN samples.

| Samples | $T_a$ (℃) ± 10℃ | Pressure during annealing (Torr) | $d$ (nm) | $T_C$ (K) (from QCC fit) | D, ($cm^2 s^{-1}$) | $R_{Max}$ ($\Omega$) | $R_N^{300K}$ ($\Omega$) | Type of MR at $T_{max}$ |
|---|---|---|---|---|---|---|---|---|
| SS1 | 750 | $1.8 \times 10^{-7}$ | 7 ± 1 | 2.45 | 0.810 | 98 | --- | |
| SS2 | 750 | $1.3 \times 10^{-7}$ | 6 ± 1 | 2.0[a] | 0.724 | 86 | 101 | *Negative* |
| SS3 | 730 | $1.5 \times 10^{-7}$ | 8 ± 1 | 2.85[a] | --- | 88 | 96 | |
| TN3 | 820 | $4.8 \times 10^{-8}$ | 8 ± 1 | 4.3 | --- | 70 | 97 | |
| TN4 | 820 | $4.7 \times 10^{-8}$ | 4 ± 0.8 | 3.55 | 0.612 | 150 | 186 | |
| TN5 | 820 | $4.6 \times 10^{-8}$ | 3 ± 0.5 | 2.93 | 0.405 | 352 | 400 | |
| TN8 | 780 | $2.2 \times 10^{-8}$ | 4 ± 0.8 | 3.1 | 0.589 | 139 | 168 | |
| TN9 | 780 | $2.1 \times 10^{-8}$ | 3 ± 0.5 | 2.43 | 0.429 | 332 | 367 | *Positive* |
| TN10 | 780 | $1.7 \times 10^{-8}$ | 2 ± 0.5 | 1.93 | 0.245 | 1246 | 1227 | |
| TN12 | 750 | $1.3 \times 10^{-8}$ | 4 ± 0.8 | 2.47 | 0.567 | 148 | 174 | |
| TN13 | 750 | $1.2 \times 10^{-8}$ | 3 ± 0.5 | 2.35 | 0.405 | 349 | 386 | |

[a] $T_c$ onset values are considered.